\documentclass[twocolumn,showpacs,preprintnumbers,amsmath,amssymb]{revtex4}

\usepackage{graphicx}
\usepackage{dcolumn}
\usepackage{bm}

\begin{document}

\preprint{APS/123-QED}

\title{
The granular silo as a continuum plastic flow:\\
the hour-glass {\it vs} the clepsydra
}

\author{L. Staron$^1$, P.-Y. Lagr\'ee$^1$ and S. Popinet$^2$}
\affiliation{%
$^1$CNRS - Universit\'e Pierre et Marie Curie Paris 6, UMR 7190, Institut Jean Le Rond d'Alembert, F-75005 Paris, France.\\
$^2$National Institute of Water and Atmospheric Research, PO Box 14-901 Kilbirnie, Wellington, New Zealand
}%

\begin{abstract}
The granular silo is one of the many interesting illustrations of the thixotropic property of granular matter: a rapid flow develops at the outlet, propagating upwards through a dense shear flow while material at the bottom corners of the container remains static.  For large enough outlets, the discharge flow is continuous; however, by contrast with the clepsydra for which the flow velocity  depends on the height of fluid left in the container, the discharge rate of granular silos is constant. Implementing a  plastic rheology in a 2D Navier-Stokes solver (following the $\mu(I)$-rheology or a constant friction), we simulate the continuum counterpart of the granular silo. Doing so, we obtain a constant flow rate during the discharge and recover the Beverloo scaling independently of the initial filling height of the silo. We show that lowering the value of the  coefficient of friction leads to a transition toward a different behavior, similar to that of a viscous fluid, and where the filling height becomes active in the discharge process.  
The pressure field shows that large enough values of the coefficient of friction ($\simeq 0.3$) allow for a low-pressure cavity to form above the outlet, and can thus explain the Beverloo scaling.  In conclusion, the difference between the discharge of a hourglass and a clepsydra seems to reside in the existence or not of a plastic yield stress.  
\end{abstract}

\pacs{45.70.-n, 83.60.Rs, 46.35.+z,83.10.Rs}%
\maketitle

\section{Introduction}

Granular matter is well-known for its ability to behave like a solid or like a fluid depending on the stress it is subjected to, transiting from one state to the other over a few grain diameters. This phenomenon is best described in terms of internal friction: a granular packing can resist shear stress below the friction threshold and remain static, but starts flowing when the friction threshold is reached.  In this respect, granular matter resembles other more classical visco-plastic materials, characterized by a yield stress and a viscosity, for instance Bingham plastics. The granular silo is one of the many interesting illustrations of this thixotropic property of granular matter: a rapid flow develops at the outlet, propagating upwards through a dense shear flow while material at the bottom corners of the container remains static \cite{janssen95,davies83,potapov96,hirshfeld97,gonzales11,djouwe12}. Beside its obvious industrial relevance, the phenomenology of the silo discharge is in itself intriguing, and raises an important interest from the scientific community. For narrow outlets, arches forms and vanish alternatively, clogging the flow with a probability depending on the outlet dimension, and leading to intermittency \cite{lepennec98,bratberg05,zuriguel05,janda09}; this regime, not accessible through continuum modeling, is not addressed in this work. For larger outlets, the flow is continuous; however, in contrast with the discharge of a Newtonian fluid for which the flow velocity  depends on the height of fluid left in the container, the discharge rate of granular silos is constant.
The independence of the discharge rate on the filling height is best demonstrated by the well-known Beverloo scaling, which relates the flow rate ($Q$) to the outlet size ($L$) \cite{beverloo61}:  in a very robust manner, experiments and discrete simulations find $Q = C \sqrt{g} (L-kd)^{N-1/2}$, where $N$ is the dimension of the problem, $d$ the grains diameter and $C$ and $k$ are constants  (we consider two-dimensional silos only in the following {\it ie} $N=2$).
 Because it implies that the velocity of the material flowing from the silo does not scale like the square root of the pressure, the Berverloo scaling is commonly accepted as the evidence of a screening effect responsible for a constant and low value of the  pressure ({\it ie} lower than the hydrostatic prediction) in the area of the outlet \cite{hilton11,perge12}. 
  {We may suppose that this is why the hourglass  was used by sailor men for navigation: the oscillations of the ship,  tilting the device and thereby changing the pressure inside, probably do not affect the discharge rate as much as it would in a clepsydra (or water-clock), making the hourglass more reliable on board.}
 The traditional physical explanation for this screening effect resorts to Janssen's analysis: the friction forces mobilized at walls reduce the apparent weight of the material in the silo and prevent the bottom area to sense the pressure, now partly sustained by the walls \cite{janssen95,ovarlez01,ovarlez03,bratberg05,sperl06}. This effect is expected to play a role only  if the width of the silo is smaller than the filling height.  Yet, the Berverloo scaling also holds for wide silos \cite{bartos06}. Moreover, the existence of a Janssen screening effect would result in the local pressure scaling like the {width} of the container, yet the Berverloo scaling involves only the outlet size. Hence, it seems very uncertain that the Janssen effect is responsible for the Berverloo scaling and the silo constant discharge rate.  This was in fact demonstrated in \cite{sheldon10} where experiments using inclined silos show that the Berverloo scaling holds for any degree of  tilt up to subvertical values. Experimental work using horizontal silos also points at a similar conclusion \cite{aguirre11}. 
On the other hand, measurements showing a dip in the value of the pressure in the area of the outlet may only reflect the fact that the existence of the outlet itself creates a low pressure boundary condition,  independently of any Janssen screening \cite{perge12}.  This does not mean however that friction is not important during the discharge of a granular silo: in this contribution, we argue that the role of friction does not involve the walls, but the bulk of the flow through the existence of a yield stress. \\
Implementing a plastic rheology (using either the $\mu(I)$-rheology \cite{jop06} or a constant friction) in a 2D Navier-Stokes solver \cite{popinet03,popinet09}, we simulate the continuum counterpart of the granular silo. Doing so, we observe a constant flow rate during the discharge and recover the Beverloo scaling independently of the initial filling height of the silo. However, we show that lowering the value of the friction lead to a transition toward a different scaling where the filling height becomes active in the discharge process. These results support the idea that the existence of a frictional yield stress can alone control the discharge of the granular silo without any Janssen effect entering into play. 
%
\section{The continuum granular silo}

\begin{figure}
\begin{minipage}{0.99\linewidth} 
\centerline{\includegraphics[width = 1.1\linewidth]{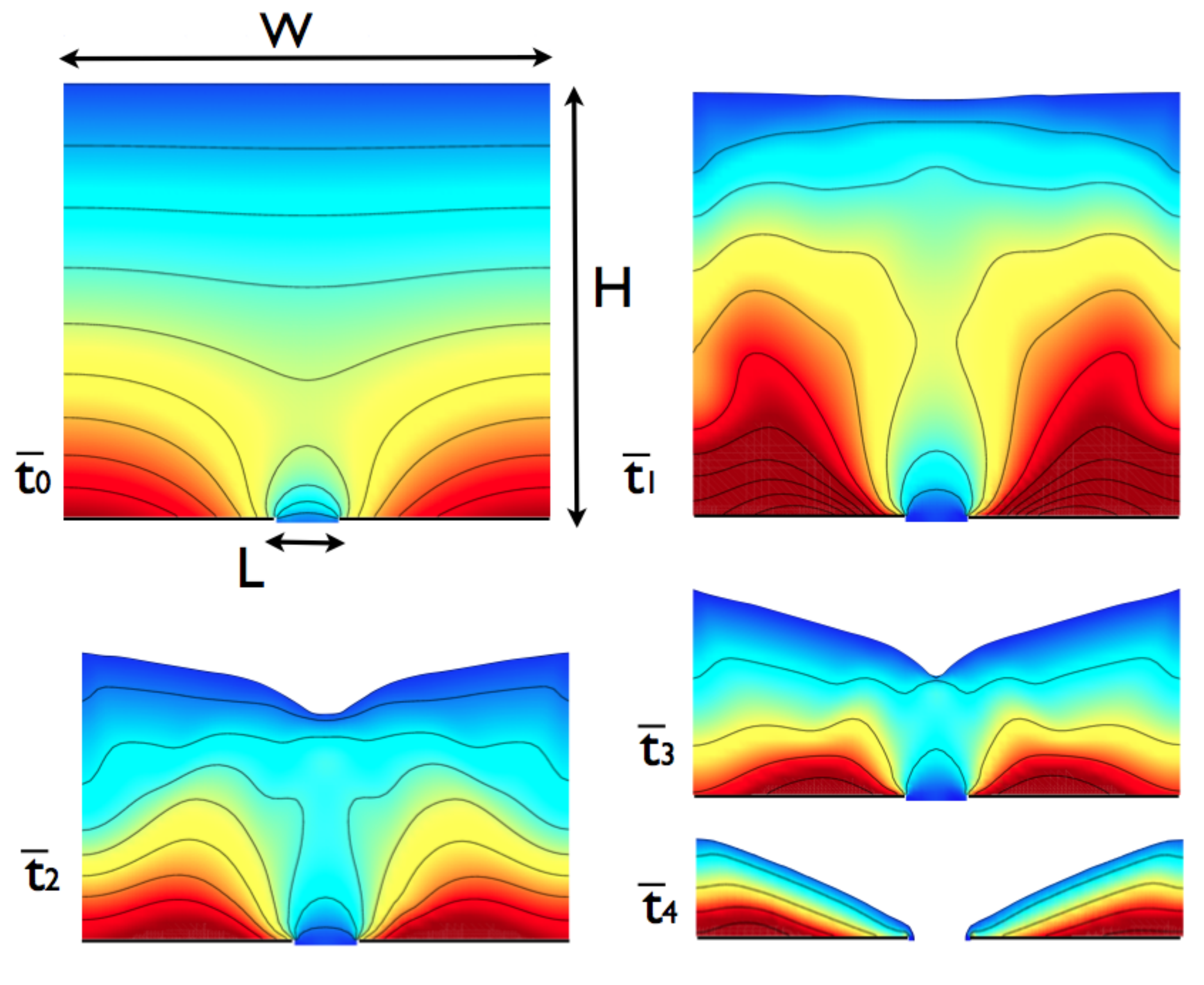}}
\end{minipage}
\caption{Pressure field during the discharge of  a plastic silo of width $W$, normalized outlet size $\bar{L}=0.125 $ and normalized filling height $\bar{H} = 0.9$ (normalized by $W$) at $\bar{t}_0 = 0$, $\bar{t}_1= 0.8$, $\bar{t}_2= 7.6$, $\bar{t}_3= 11.5$, and at $\bar{t}_4$ the final state (normalized by $\sqrt{W/g}$). The color scale varies from one picture to the other to ensure maximum contrast: the highest bound (red color) is set to $\bar{P} = 0.6$ for $\bar{t}_0$, $\bar{t}_1$ and $\bar{t}_2$,  to $\bar{P}=0.36$  for $\bar{t}_3$, and to $\bar{P}= 0.12$ for $\bar{t}_4$ (normalized by $\rho g W$). }
\label{pression}
\end{figure}

The simulations were performed using the Gerris flow solver in two dimensions, which solves  the Navier-Stokes equation for a bi-phasic mixture applying a Volume-Of-Fluid approach \cite{popinet03,popinet09}. The existence of two fluids translates numerically  in different properties (viscosity and density) on the simulation grid following the advection of the volume fraction representing the proportion of each fluid. In our case, one fluid stands for granular matter (characterized by the coefficient of internal friction $\mu$) and the other stands for the surrounding air (with a lower density and lower viscosity, see \cite{lagree11} for details); the position of the interface between the two is solved in the course of time based on the spatial distribution of their volume fraction. 
The viscosity  $\eta$ of the granular matter is defined by mean of the friction properties \cite{jop06}: 
\begin{equation}
\eta = \min\left(\frac{\mu P}{D_2}, \eta_{max}\right), 
\end{equation}
where $\mu$ is the effective coefficient of friction of the granular flow, $P$ is the local pressure and $D_2$ is  the second invariant of the strain rate tensor $\boldsymbol D$: $D_2=\sqrt{D_{ij}D_{ij}}$. For large values of $D_2$, the viscosity is finite and proportional to $\mu$ and P; when $D_2$ reaches low values, the viscosity $\eta$ diverges. Numerically, this divergence is bounded by a maximum value $\eta_\text{max}$ chosen to be $10^4$ times the minimum value of $\eta$; we have checked that the choice of $\eta_\text{max}$  did not affect the results as long as $\eta_\text{max}$  is large enough. \\
{Based on earlier work on continuum modeling of rapid non-uniform granular flows showing the better performances of  the $\mu(I)$-rheology compared to constant friction \cite{lagree11},  the effective friction properties $\mu$ of the granular continuum is calculated using the following dependence}: $\mu$ is a function of the non-dimensional number $I = d D_2/\sqrt{P/\rho}$, where $d$ is the mean grain diameter and $\rho$ the density ($W=90d$ in the following, with $W$ the silo's width) through the relation
$\mu = \mu_s + \frac{\Delta \mu} {1+ I_0/I }$
 where $\mu_s$, $\Delta \mu$ and $I_0$ are constants \cite{jop06,lagree11}.  
{Following \cite{lagree11}, $\Delta \mu$ and $I_0$ were set respectively to $0.28$ and $0.4$ and not varied. The value of  the static coefficient of friction $\mu_s$ was set to $0.32$, $0.2$ and $0.1$ in order to study its role in the silo discharge.  Moreover, simulations with a constant friction model, {\it ie } without dependence on $I$, were also performed (section \ref{friction}).}  \\ 
 The silo is flat-bottomed, of width $W$, filling height $H$ and with an outlet of size $L$ (see Figure \ref{pression}).  The width $W$ is divided in 64 computation cells in the bulk, refined to 256 at the bottom, so that the outlet is defined using 16 to 72 computation cells. In the forthcoming analysis, all quantities are normalized as follows:  silo height  $\bar{H} = H/W$, outlet size  $\bar{L} = L/W$, volume  of material left in the silo $\bar{V} = V/W^2$,  flow rate $\bar{Q}= Q/ W\sqrt{gW}$, time $\bar{t}= t/\sqrt{W/g}$ and  pressure  $\bar{P} = P/\rho g W$.\\
   A no-slip boundary condition is imposed at the side-walls and at the bottom-wall; 
   {additional simulations with a free-slip boundary condition at the side-walls show that the aspects discussed in this paper remain unchanged}. A zero pressure condition is imposed at the top-wall and at the outlet.
\begin{figure}
\begin{minipage}{0.99\linewidth} 
\centerline{\includegraphics[width = \linewidth]{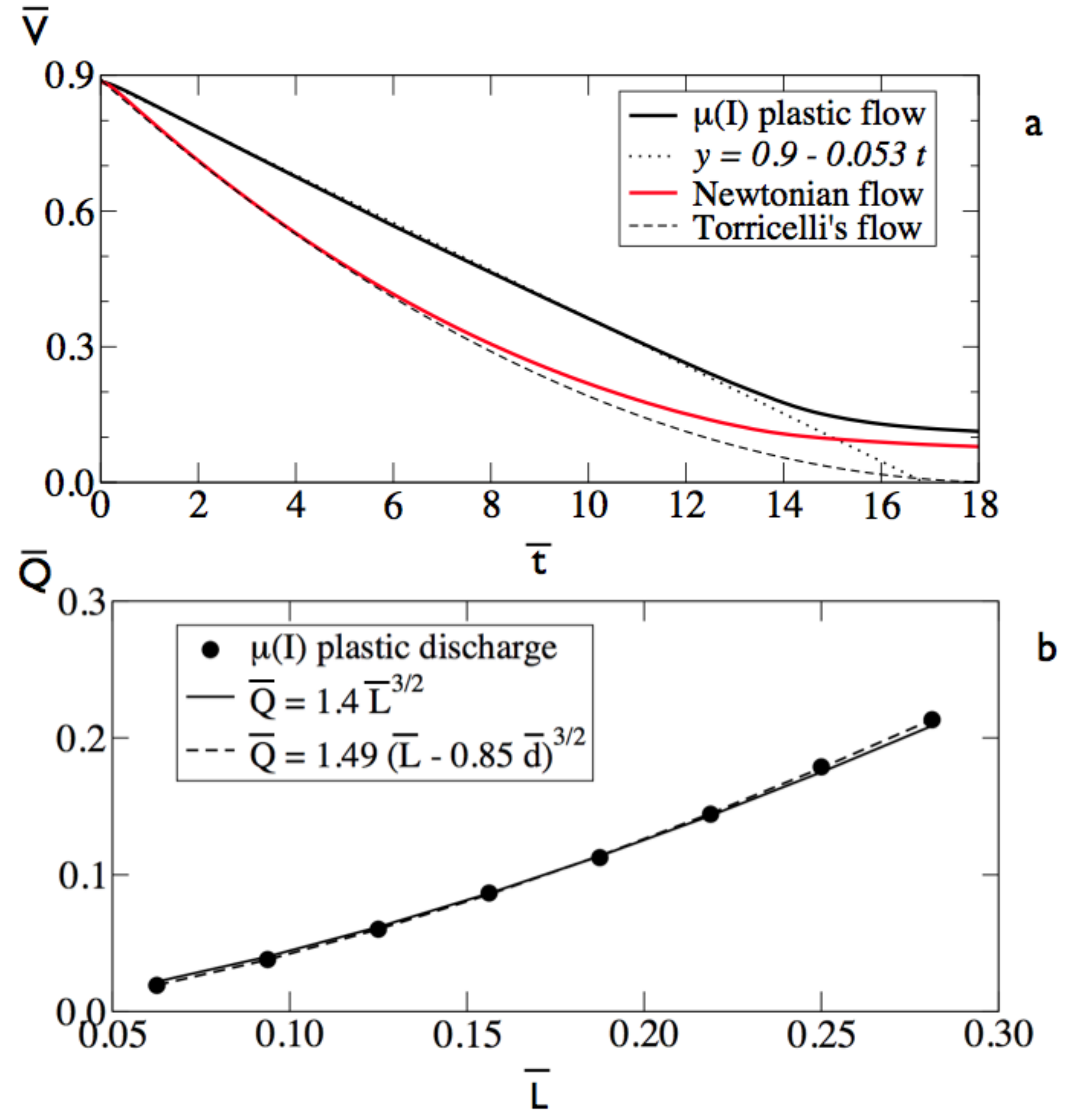}}
\end{minipage}
\caption{ a- Volume $\bar{V}$ of material remaining in the silo  as a function of  time for  an outlet size $\bar{L}=0.125$ and filling height $\bar{H} = 0.9$ in the case of a plastic flow ($\mu_s=0.32$, $\Delta \mu=0.28$ and $I_0=0.40$) and in the case of a Newtonian flow ($\eta= 0.01 \rho g^{\frac12}W^\frac32$); b- Berveloo scaling obtained for $\bar{L}$ varying between $0.0625$ and $ 0.28125$.}
\label{discharge1}
\end{figure}

\section{A constant discharge rate}
Figure \ref{pression} shows the time evolution of a continuum granular silo of  initial filling height $\bar{H}= 0.9$, and  outlet size $\bar{L}= 0.125$; the static friction is set to $\mu_s=0.32$ (with $\Delta \mu=0.28$ and $I_0=0.40$). The color scale represents the pressure field. We observe that the pressure field strongly differs from what would be expected in the hydrostatic case, and is non-uniform in the transverse direction.  The region above the outlet coincides with a low pressure cavity surrounded by two high-pressure dome-like areas: {the pressure jump at the outlet overcomes the frictional yield stress, creating  a large shear and a low viscosity area. Meanwhile, pressure gradients decrease in the bulk, and a highly viscous mass forms above the outlet. The fact that the yield stress is frictional, and depends on the pressure, implies that it readapts throughout the discharge, thereby probably allowing the cavity to survive  until the end of it.} When the material left in the silo stops flowing, it remains at equilibrium with a shape depending on the yield stress, {\it ie} depending on the coefficient of friction $\mu$. \\
Figure \ref{discharge1}-a shows the volume of material remaining in the silo in the course of time for the same system. We observe a linear evolution throughout the discharge for the granular fluid, revealing a constant flow rate as in real discrete granular silos. We measure the value of the viscosity in the vicinity of the outlet and find a roughly constant value during the discharge equal to $\bar{\eta} = 0.01$ (normalized by $\rho g^{\frac12}W^\frac32$). Using this value for the viscosity, we simulate the discharge of a silo filled with Newtonian fluid; the corresponding volume-{\it vs}-time evolution  is shown in Figure \ref{discharge1}-a, and is non-linear, suggesting a dependence on the height of material left in the silo, as in the case of an hydrostatic pressure field.  For comparison, we plot the solution of the (non-dimensional) Torricelli discharge for an ideal fluid: 
\begin{eqnarray}
\nonumber
\frac{d \bar{h}}{d\bar{t}} &=& - \bar{L}\sqrt{2\bar{h}},\\ \nonumber
\bar{h}(\bar{t}) & = & \left(\sqrt{\bar{H}} - \frac{\bar{L}}{\sqrt{2} } \bar{t}\right)^2,\nonumber
\end{eqnarray}
 where $\bar{h}$ is the instantaneous height of material remaining in the silo (normalized by $W$) at time $\bar{t}$ (normalized by $\sqrt{W/g}$). Torricelli's discharge matches the onset of the discharge of the Newtonian  fluid provided we chose a smaller numerical value for $\bar{L}$ than that of the simulation ($\bar{L}=0.0714$ instead of $0.125$). Comparing the discharge of the Newtonian fluid and the granular plastic fluid  thus points at  the plastic property of the flow as responsible for the constant nature of the discharge rate in the latter case. \\
  We observe that the flow rate remains constant throughout the discharge of the plastic silo over a large range of outlet size $\bar{L}$. Varying $\bar{L}$ between $0.0625$ ({\it ie} 16 computation cells) and $ 0.28125$ ({\it ie} 72 computation cells), we measure the flow rate $\bar{Q}$, and search for a relation satisfying the shape of the Beverloo scaling: 
  \begin{equation}
 \bar{Q} = C \left({\bar{L} - k }\right)^{\frac32}
 \label{Bev1}
\end{equation}
where C and k are constants.  For a continuum silo, the numerical value of $k$ is expected to be zero, in contrast to granular silos where the grain diameter imposes a volume of exclusion reducing the effective size of the outlet. Imposing $k=0$, we recover the Berverloo scaling  with a good accuracy, giving $C=1.4$ (see Figure \ref{discharge1}-b). Note however that making no assumption on the fitting parameters, the best fit gives $k=0.00938=0.85\bar{d}$, where $\bar{d}=1/90$ is the  grain diameter used in the $\mu(I)$-rheology. {Although the value of $k$ is not completely negligible, we consider nevertheless that it is not physically significant.}%

\begin{figure}
\begin{minipage}{0.99\linewidth} 
\centerline{\includegraphics[width = \linewidth]{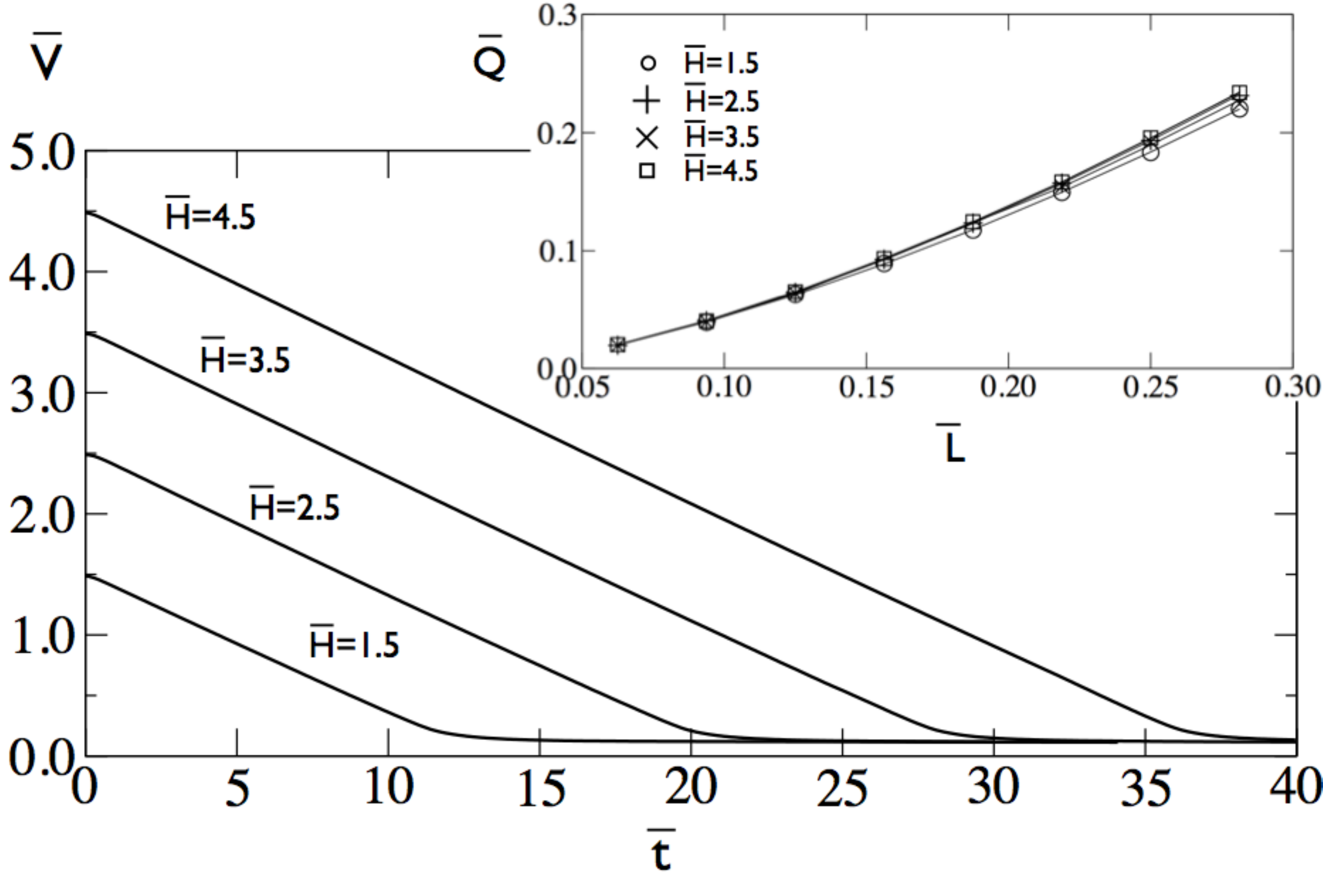}}
\end{minipage}
\caption{ Volume $\bar{V}$ of material remaining in the silo as a function of  time for  $\bar{L}=0.1875$ and filling heights $\bar{H} = 1.5$, $\bar{H} = 2.5$, $\bar{H} = 3.5$ and  $\bar{H} = 4.5$ for a  $\mu(I)$ plastic flow  ($\mu_s=0.32$, $\Delta \mu=0.28$ and $I_0=0.40$). Inset: Berverloo scalings corresponding to each case.}
\label{discharge2}
\end{figure}
\begin{figure}
\begin{minipage}{0.99\linewidth} 
\centerline{\includegraphics[width = \linewidth]{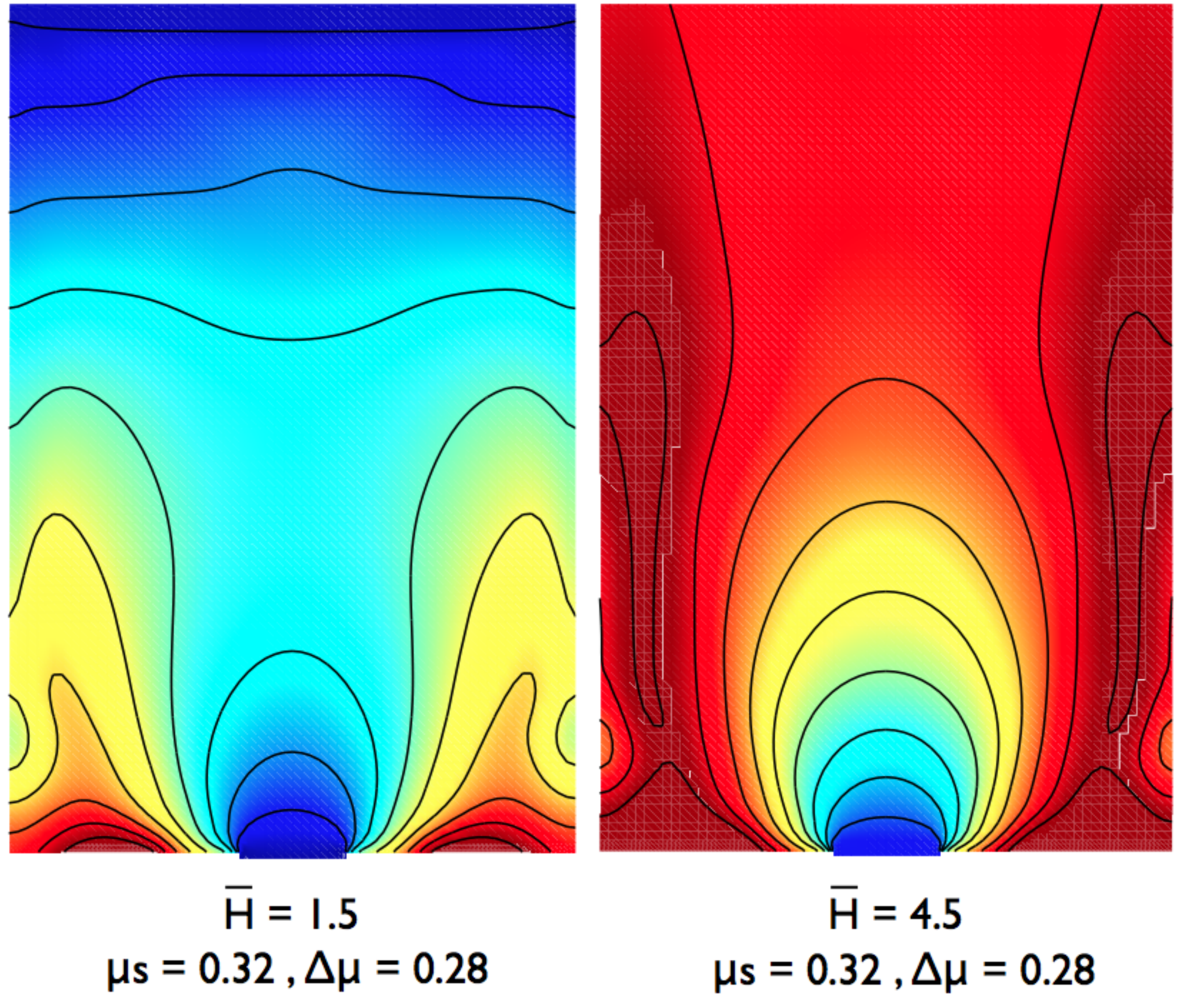}}
\end{minipage}
\caption{Pressure field in the early stage of the discharge for two granular plastic silos of  initial filling heights $\bar{H} = 1.5$ (right)  and  $\bar{H} = 4.5$ (left), outlet $\bar{L}=0.1875$  and $I$-dependent friction coefficient $\mu_s=0.32$ ($\Delta \mu=0.28$ and $I_0=0.40$).  The color scale is identical on both images: the highest bound (red color) is set to $\bar{P}=1.3$; the pressure jump between two isolines is $0.15$. }
\label{Pression2}
\end{figure}

\section{Increasing the filling height}
To check whether the discharge rate remains constant irrespective of the initial filling height, we perform additional simulations with $\bar{H}= 1.5$, $2.5$, $3.5$ and $4.5$, with the same rheological parameters as previously ($\mu_s=0.32$, $\Delta \mu=0.28$ and $I_0=0.40$). Figure \ref{discharge2} shows the volume $\bar{V}$ left in the silo in the course of time for an outlet size $\bar{L}=0.1875$ for all four cases. We observe a linear evolution of very similar slope, suggesting that the discharge rate is essentially constant and independent of the filling height. However, closer inspection shows that the slope varies slightly during the discharge: for  $\bar{H} = 4.5$ for instance, the initial flow rate has decreased of $1.4 \%$ halfway through the discharge. Moreover, a slight increase of the flow rate is observed for larger filling height: approximating the discharge by an affine function over its full duration, we find a flow rate increase of $1.8\%$ for $\bar{H} = 2.5$  and an increase of $3.9\%$ for  $\bar{H} = 4.5$ compared to the case of $\bar{H} = 1.5$.\\
 Measuring the flow rate $\bar{Q}$ in the early stage of the discharge for the different values of $\bar{H}$ and different outlet dimension $\bar{L}$, we recover the Berveloo scaling (\ref{Bev1}), but with coefficients varying slightly with the value of $\bar{H}$ (Figure  \ref{discharge2}, inset). This weak influence of the initial filling height is maximum in the early stage of the discharge, but vanishes at the end: the curves shown in Figure  \ref{discharge2} can be superimposed when shifted towards the final stage of the discharge.\\
 %
Altogether, the flow rate during the discharge of continuum granular material is very weakly affected by the initial filling height $\bar{H}$.  By contrast, the pressure field strongly changes with the value of $\bar{H}$, as is visible on the two snapshots shown in Figure \ref{Pression2} for $\bar{H}=1.5$ and $\bar{H}=4.5$ in the early stage of the discharge. In both cases however, we observe a low pressure cavity in the vicinity of the outlet, suggesting that the discharge is affected by this local pressure condition and insensitive to the mean pressure in the silo. \\
\section{Influence of the internal friction}
\label{friction}
\begin{figure}
\begin{minipage}{1.0\linewidth} 
\centerline{\includegraphics[width = \linewidth]{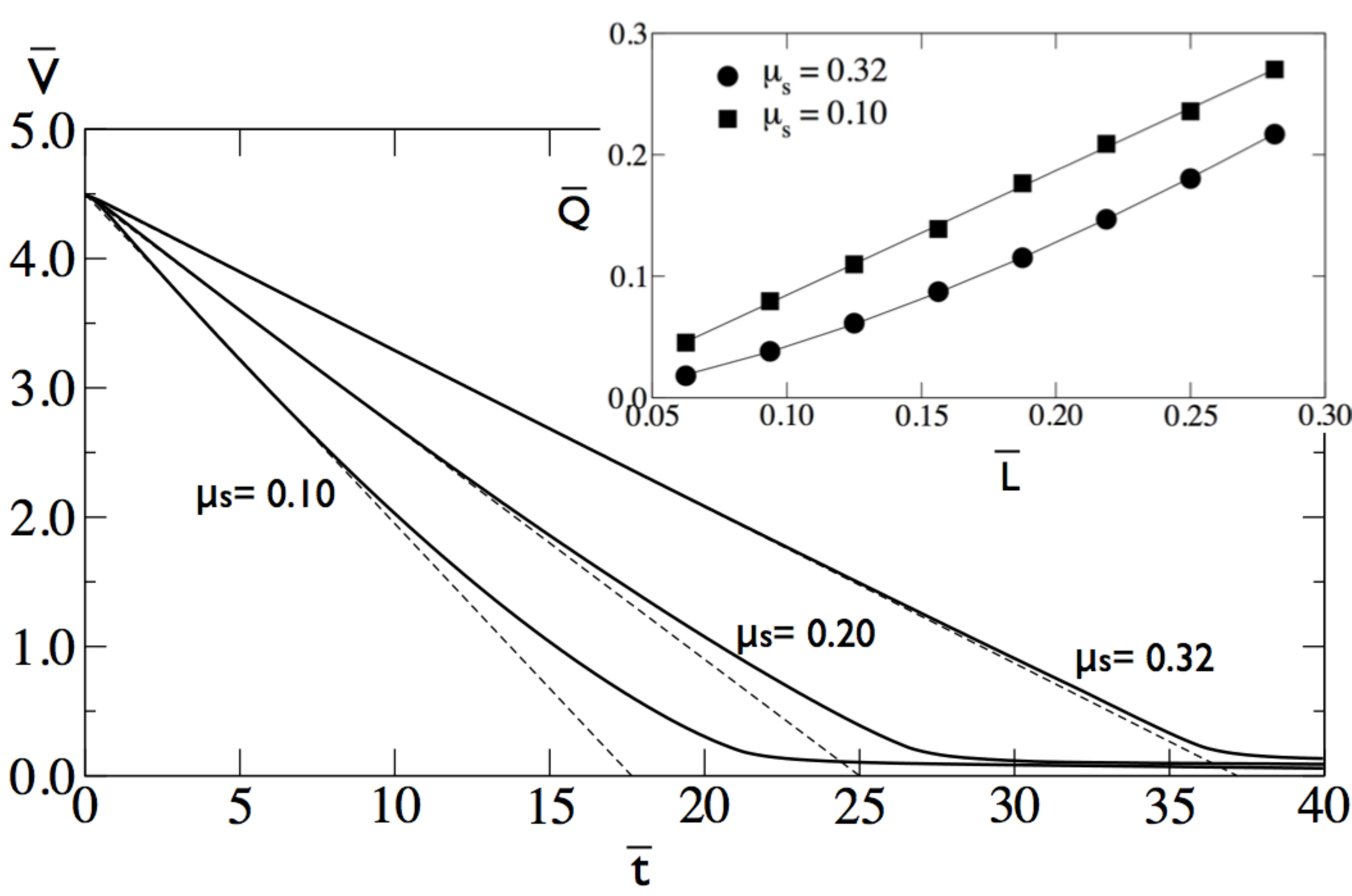}}
\end{minipage}
\caption{ Volume of material  $\bar{V}$ remaining in the silo as a function of  time   $\bar{t}$ for $\bar{L}=0.1875$ and filling height $\bar{H} = 4.5$,  for a  $\mu(I)$ plastic flow with $\mu_s=0.10$, $\mu_s=0.20$, and  $\mu_s=0.32$ ($\Delta \mu=0.28$ and $I_0=0.40$). Inset: Flow rate $\bar{Q}$ as a function of outlet size $\bar{L}$  for $\mu_s=0.10$ and  $\mu_s=0.32$.}
\label{discharge3}
\end{figure}
\begin{figure}
\begin{minipage}{0.99\linewidth} 
\centerline{\includegraphics[width = \linewidth]{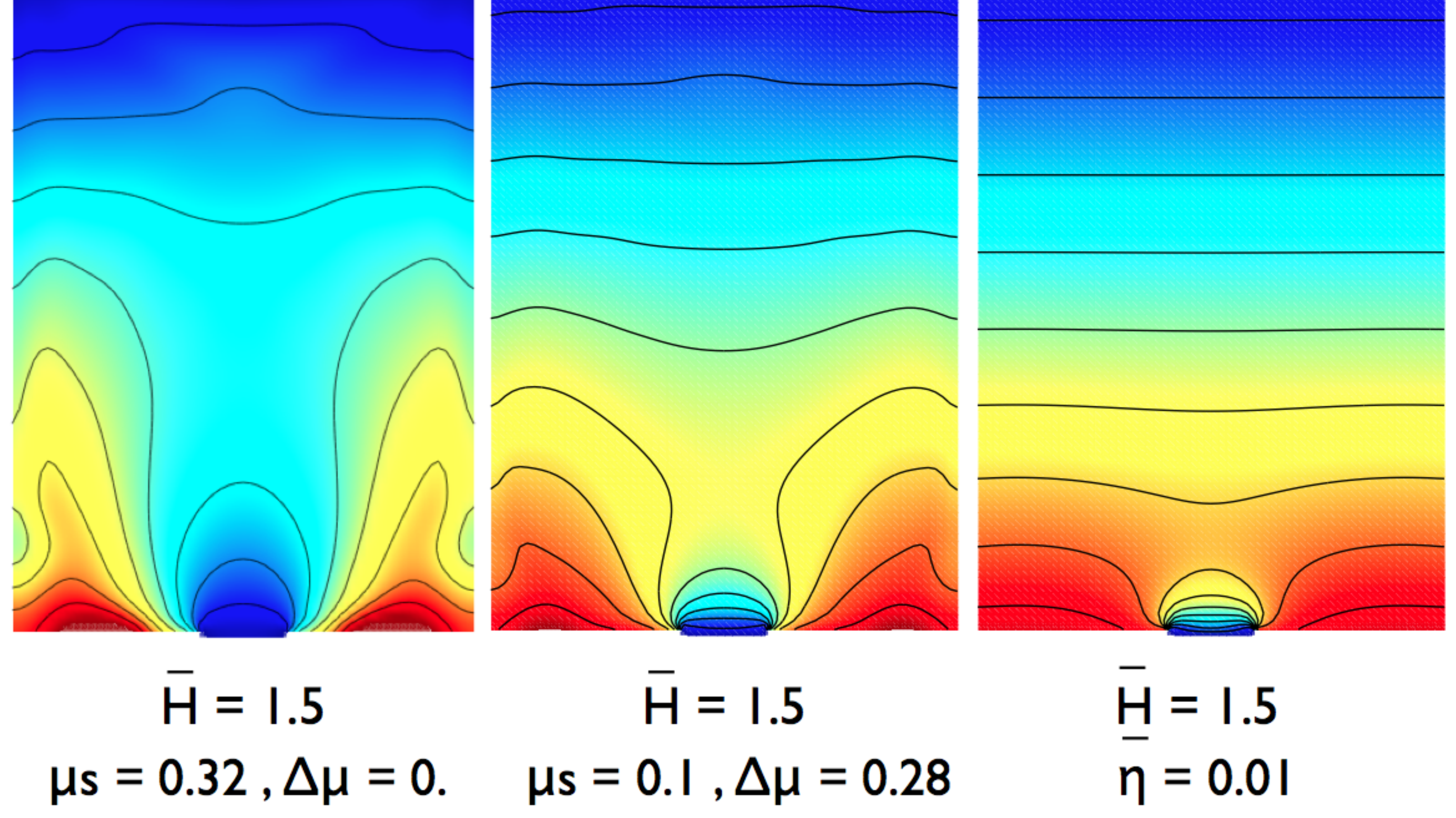}}
\end{minipage}
\caption{Pressure field in the early stage of the discharge for a granular plastic flow with  a constant friction coefficient  $\mu_s=0.32$  ($\Delta \mu=0.$, left),  a $I$-dependent friction $\mu_s=0.10$  ($\Delta \mu=0.28$ and $I_0=0.40$, center) and for a Newtonian flow of viscosity $\bar{\eta}=0.01$ (normalized by $\rho g^\frac12 W^\frac23$) (left). Outlet size $\bar{L}= 0.1875$, filling height $\bar{H} = 1.5$.  The color scale is identical on all images and identical to Figure \ref{Pression2}: the highest bound (red color) is set to $\bar{P}=1.3$; the pressure jump between two isolines is $0.15$. }
\label{Pression3}
\end{figure}

We suspect the deviation of the discharge rate from the hydrostatic case to be the result of the non-newtonian nature of the $\mu(I)$-rheology, and more specifically of the existence of a yield stress; accordingly, we expect the value of the coefficient of friction to have important repercussions on the discharge flow.  Figure \ref{discharge3} shows the discharge of a silo with $\bar{H}=4.5$ and outlet $\bar{L}=0.1875$ for different values of the coefficient of static friction $\mu_s=0.10$, $\mu_s=0.20$ and $\mu_s=0.32$: we observe that a smaller friction coefficient induces an earlier departure from the linear evolution observed for larger friction. Since friction does not only act upon the yield stress, but also upon the value of the viscosity, it affects the discharge duration too.\\
 For the three values of the static coefficient of friction, and for a silo with $\bar{H}=1.5$, we vary the outlet size $\bar{L}$ between  $0.0625$ and $0.28125$ and measure the flow rate $\bar{Q}$. For $\mu_s=0.10$ and $\mu_s=0.20$, the flow rate is not constant all through the discharge. Hence, we consider the early stage of the discharge only, where an affine regression seems reasonable. As already seen in previous sections, the flow rate in the case $\mu_s=0.32$ obeys the Beverloo scaling (see Figure \ref{discharge1});  in the case of lower friction however ($\mu_s=0.10$), the Beverloo scaling is no more relevant, and the flow rate increases linearly with the outlet size $\bar{L}$ (Figure \ref{discharge3}, inset), thus making a dependence on $\bar{H}^{1/2}$  dimensionally possible.\\
 {An other aspect is the importance of the $I$ dependence on the friction model. Although fully addressing this question implies detailed comparison with discrete systems (either experimental or numerical), we can nevertheless check how the height independence of the silo discharge is affected by suppressing this dependence and set $\Delta \mu=0$. Figure \ref{Pression3} shows the pressure field in a silo with $\mu_s=0.32$, $\Delta \mu=0$ and $\bar{H} =1.5$ (to compare with Figure \ref{Pression2}): we observe no significant difference with the case $\Delta \mu \neq 0$.  Plotting the discharge rate against outlet size allows for the recovery of the Beverloo scaling as well (not shown). It is thus clear that the existence of a frictional threshold alone is enough for reproducing the height independence, without the need of a $I$ dependence. The effect of the latter being to increase the friction close the the outlet, it may become important when comparing quantitatively discrete and continuum granular silo discharge rate. This discussion is however beyond the scope of this paper.\\ 
 Figure \ref{Pression3}  also presents the pressure field in a plastic silo with weak friction $\mu_s=0.1$ and in a Newtonian silo of viscosity $\bar{\eta}=0.01$ (normalized by $\rho g^\frac12 W^\frac23$), and shows that the existence of the low pressure cavity is contingent on the existence of a yield stress  induced by sufficiently large friction (Figure \ref{Pression3}).}

\section{Discussion}

Granular matter forms a specific class of plastic fluids for which yield stress and  viscosity are not independent, but related through friction properties. Hence discriminating between the respective roles of yield stress and viscosity is difficult. However, the continuum simulations presented in this paper show that {\it i)} a friction-dependent viscosity  (either $\mu(I)$ or constant friction) leads to a constant flow rate as is the case for real granular silo and  {\it ii)} decreasing the coefficient of friction leads to a dependence on the height of material remaining in the silo, as is the case for Newtonian fluids.  Decreasing the coefficient of friction to a value as small as $0.1$  may not be physical for granular matter; however, this extrapolation tends to show that the constant shear rate results from the existence of a yield stress. 
 {For sufficiently large values of the coefficient of friction ({\it ie} $\geq 0.3$), the pressure field shows the existence of a low pressure cavity in the vicinity of the outlet, despite high values of the pressure elsewhere in the silo (Figure \ref{Pression2}).  The fact that the yield stress ($\mu P$) depends on the mean pressure in the system is certainly crucial in the ability of the cavity to remain identical all though the discharge.}
 This cavity does not exist during the discharge of a Newtonian flow, and is very reduced in the case of a plastic flow with a small friction coefficient $\mu_s=0.10$ (Figure \ref{Pression3}).
{ It thus seems reasonable to conclude that the frictional properties of the continuum granular material are at the origin of the low-pressure cavity forming above the outlet; this coincides with a high-shear-low-viscosity region resembling the free fall arch described in granular systems \cite{hilton11}.} The fact  that the flow is controlled by the local conditions close to the outlet rules out the Janssen effect as explanation for the Beverloo scaling (as did before \cite{sheldon10,aguirre11} experimentally). In conclusion, the difference  between the discharge of an hourglass and a clepsydra seems to reside in the existence or not of a frictional yield stress.  \\ 
{The ability of the continuum $\mu(I)$-rheology to reproduce the behavior of granular systems when implemented  in a Navier-Stokes solver was discussed in \cite{lagree11}. In this contribution, we show the ability of the same model to reproduce the silo phenomenology, without discussing in depth the relevance of the $I$-dependence. It should be noted that a constant friction model leads to the silo phenomenology too, as far as height independence and Beverloo scaling general shape are concerned.}  More detailed aspects like comparison of velocity and pressure profiles, inner deformations and surface shape are beyond the scope of the present work.  They are the subject of a forthcoming  dedicated communication.


%
%
%

\end{document}